\documentstyle{mn}

\title[Galaxies in the Hubble Deep Field]
{The nature of the faint galaxies in the Hubble Deep Field}

\author[B. Mobasher et al.]
{B.~Mobasher, M.~Rowan-Robinson, A.~Georgakakis, N.~Eaton\\
Astrophysics Group, Blackett Laboratory, 
Imperial College, Prince Consort Rd, London SW7 2BZ}

\begin{document}

\maketitle
\begin{abstract}
We present a study of the galaxies found in the Hubble Deep Field.
A high proportion of HDF galaxies are undergoing a strong episode of star
formation, as evidenced by their very blue colours.  A wide range of
morphological types is found, with a high proportion of peculiar and merger
morphologies.

Fitting the multiband spectra with redshifted SEDs of galaxy types E to HII, 
we predict the spectral types and redshifts of galaxies detected in the HDF. 
We find a median redshift of 1.6, with $68\%$ having $z > 1$ and $31\%$ with
$z >2$. The I-band absolute magnitude distributions as a function of galaxy
types show a plausible trend of decreasing luminosity towards later 
types. 
The derived I-band luminosity function agrees well with that from
the Canada-France survey (Lilly et al 1996) for $z < 1$ and shows strong
luminosity evolution at $M_I < -21$ for $1 < z < 3$, comparable to the rate 
seen in quasars and starburst galaxies.

We have predicted infrared and submillimetre fluxes assuming most of the
galaxies are undergoing a strong starburst.  Several planned space-borne and
ground-based deep surveys are capable of detecting interesting numbers of HDF
galaxies.
\end{abstract}


\section{Introduction}
 In December 1995 an area of five square arcminutes at RA 12h 36m 49.4s
Dec +62 12 58 (J2000) was surveyed by the Hubble Space Telescope to an
unprecedented depth (Williams et al 1996). This survey, refered to as the
Hubble Deep Field (HDF), consists of three wide field areas, each of size
75 arcsec square, carried out by the WFPC2, and a smaller area of 35 arcsec
square covered by the planetary camera. The HDF is over one magnitude deeper
than the deepest surveys previously done by the HST and is carried out in
four bands; F300W(300nm), F450W(450nm), F606W(600nm) and F814W(800nm),
roughly corresponding to the UBVI system.  The field is chosen to have a
representative number density for galaxies and be devoid of bright stars.

One of the main applications of the HDF is to study the nature of the faint
population of galaxies found in ground based optical surveys.  Recent models,
interpreting deep counts of galaxies, have invoked bursts of star formation,
merging of galaxies (Broadhurst et al 1988), dust obscuration
(Franceshini et al 1994), the presence of a new population of dwarf galaxies
(Cowie et al 1990) and a steep faint-end slope for the local luminosity
function (Koo et al 1993).  The degree to which we can discriminate between
these competing scenarios is limited by the size and depth of the available
surveys, the morphological type information obtained from them at deep levels
and the wavelength in which they have been carried out. The HDF and its
follow up spectroscopic observations are expected to disentangle some of
these  scenarios and provide a natural extension to ground-based optical
surveys (Smail et al 1995, Lilly et al 1996).  Recently, Abraham et al (1996)
performed morphological type classification of the HDF galaxies with $I < 25$
mag and conclude that at this depth, the classical Hubble system fails to
explain the large fraction of peculiar/Irregular/merger galaxies observed in
the HDF.

In this study, we explore the nature of the faint galaxy population observed
in the HDF. Using the available multi-waveband information, we predict
photometric redshifts  and spectral types of the galaxies detected in the HDF.
These are used to make a preliminary study of the distribution of the faint
blue population in redshift and luminosity space.  A promising aspect of
future studies of the HDF is follow-up at longer infrared and sub-millimetre
wavelengths.  We make predictions of the fluxes expected at these wavelengths,
assuming most of the galaxies are undergoing a major episode of star formation.
We assume $H_0=50$ km/sec/Mpc and $q_0=0.5$ throughout this paper.

The catalogue generation is discussed in the next section. Section 3 presents
a study of the faint blue galaxy population in the HDF, followed by estimates
of photometric redshifts and spectral types of the galaxies in section 4.
The Hubble diagram and luminosity function of the HDF galaxies are presented in
section 5, and prediction of their infrared and submillimetre fluxes in
section 6. Conclusions are summarised in section 7. 

\section{Creation of a galaxy Catalogue}
Independent galaxy catalogues were first generated in all the four bands.
Object identifications and star/galaxy separations were performed, using the
PISA software in the STARLINK environment. This was further confirmed using
the FOCAS/IRAF source identification package. Aperture photometry was then
carried out with PHOTOM on all the identified galaxies, using an aperture
size of $0.5''$ diameter. 
For $0.5<z<3$, this corresponds to a linear diameter of 
$4.0 \pm 0.2\ (h/50)^{-1}$ kpc. The photometry zero points provided by 
the STScI were used. 
The catalogues for different bands were cross correlated and objects
with detections in more than one band were identified. A total of 1761
galaxies were identified in the 3 WFPC2 fields, of which 1536 were detected in
the I-band. 327 galaxies have been detected in all  four bands, 577 in the
3 bands (IVB), 397 in the 2 bands (IV), while 230 were detected in one band 
only (the I band in the case of 229 galaxies).  A total of 1307 galaxies were
detected in both I and V bands (and in other bands for 910 objects). This
sample forms the basis of the present study, ie. we ignore the 230
single band detections. The magnitudes at which serious
incompleteness sets in are I$\sim$ 28, V $\sim$ 28.5, B $\sim$ 28.5, 
U $\sim$ 26, with the
completeness limit for statistical purposes being about 0.5 magnitudes
brighter than these values. 

Morphological type classification of galaxies detected in four bands was
carried out by two of us (AG and BM). Galaxies were divided into four
categories; ellipticals, spirals, irregular/peculiar and mergers, the latter
being defined as objects with multiple nuclei. 

\section{The faint blue galaxy population in the HDF}
The (B-V)-(U-B) colour-colour diagram for different morphological types of
galaxies, detected in all 4 bands, are compared with model predictions in
Fig. 1. The synthetic colours are predicted at different redshifts using the
observed SEDs of elliptical, spiral (Sbc) and HII galaxies.  The SEDs of
elliptical and spiral galaxies are taken from Yoshii and Takahara (1986) and
that for the HII galaxies is from the observed continuum of Tol 1924-416
(Calzetti and Kinney (1992). The study of the UV continua of HII galaxies by
Kinney (1993) indicates that the UV spectrum of this galaxy is typical of
HII galaxies. The predicted colours for $z > 1.5$ are based on extrapolation of
the observed/model continua to the Lyman limit, beyond which we assume the
continua to drop steeply. The predicted colours are calculated in the HST
filters.

The striking feature in Fig. 1 is the presence of two different branches with
similar U-B but different B-V colours.  The irregular/peculiar galaxies lie
at the blue end of the colour-colour diagram with the merging galaxies
extending to the blue branch, presumably due to star formation induced by
galaxy mergers. 

The range of predicted colours agree with the observed data.  The blue branch
is reasonably well fitted by the elliptical and HII galaxy models, 
shifted to high redshifts ($z \sim 1.5-2$) for the very blue galaxies. At
these redshifts, the $U-B$ colours rapidly increase as the Lyman limit
comes into the U-band. 
Also, the spiral galaxy locus and the red part of the HII galaxy model 
are in fairly good agreement with the observed data in the range $z=0-0.5$.

\section{Photometric redshift estimates}
Since most of the HDF galaxies are too faint for spectroscopic
redshift measurements with even the largest ground-based telescopes, 
it is useful to estimate their photometric redshifts, using the available 
multi-wavelength data. 
The existence of the U-band data, covering the 4000 A break is crucial for
such purpose (Connolly et al 1995; Koo 1986).

The observed, rest frame SEDs for 6 different types of galaxies
(E/SO, Sab,Sbc,Scd,Sdm from Yoshii and Takahara 1986 and HII from Calzetti and Kinney
1993, as above) are used to produce a grid of SEDs in the log(1+z) range 0.01
to 0.6 in 0.01 intervals. These are then compared with the observed SEDs for
the HDF galaxies and the best fit was selected by  least squares. The
corresponding redshift and spectral type was then associated with that galaxy.
The agreement between the spectral and morphological type classification was
found to be better than 75 percent.  We find that the number of galaxies of
SED-type E/SO, Sab, Sbc, Scd, Sdm and HII are 219, 193, 81, 169, 285, 366
respectively.  

The uncertainty in the redshift can be estimated by calculating the value of
$\chi^2$ as a function of z for each galaxy type.  
For an assumed magnitude error
of ± 0.06 mag. in each band, the typical uncertainty in (1+z) ranges from
$3 - 10\%$ for a galaxy detected in all 4 bands, with only a minority giving
acceptable solutions for more than one galaxy type. However, for objects 
detected in
only 3 or 2 bands, the uncertainty in redshift increases and there is often
more than one acceptable redshift/galaxy type solution.  The redshift
distribution for the whole sample is shown (as a function of galaxy type) in
Fig 2 .  No obvious bias is present and on the whole, there is smooth
distribution of redshifts for each galaxy type.  The strong peaks at
$z = 2$ and $z = 3$ are probably artefacts of the phometric redshift method
and a result of our ignorance of the far-UV properties of galaxies . 
To explore the sensitivity of our results to selection effects and redshift
errors, we also consider the redshift distributions for galaxies detected in
all the 4 bands (Fig.2- dotted line). These galaxies, which have more
accurate redshifts, follow the redshift distribution, based on all the objects,
producing again a peak at $z\sim 2$. 

There are 109 objects which are detected in B, V, I bands but not in
the U-band, and for which the U-band limit ($U < 26.5$) 
implies $f_\lambda (U) < f_\lambda (B)$. 
For these, we set $U=26.5$ mag., resulting to a redshift in the range
1.9--2.2 and for many of these cases, to a much higher redshift, up to 3.5. 
Also, there are 9 galaxies detected only in the I and V bands, with blue (V-I)
colours, but for which the non-detection in the U- and B-bands implies a
significant drop in the observed blue continuum.  These are candidates for
the Lyman limit being redshifted into the blue band, so that they would be
expected to have z $\geq$ 3.6. These are obvious candidates for future
spectroscopic observations.

The I-band absolute magnitude distribution $M_I$ for each galaxy type are
presented in Fig 3.  These show a highly plausible trend of decreasing
absolute magnitude towards later type galaxies. The absolute magnitude
distribution, based on galaxies with more accurate redshifts (ie. those
detected in all the 4-bands), are also presented (Fig. 3-dotted line)
and show a similar trend with type. This indicates that the increase in
errors in redshifts for some of the objects here does not affect the main 
results of this study. 

\begin{table*}
\begin{minipage}{130mm}
\caption{Predicted cumulative numbers of HDF galaxies as a function of
infrared and submillimetre flux-densities}

\begin{tabular}{lrrrrrrrr}
wavelength: & 6.7 & 15 & 60 & 90 & 200 & 400 & 800 & 1200 $\mu m$ \\
\\
flux-density: \\
10 mJy	 & 0 & 0 & 0 & 0 & 4 & 1 & 0 & 0 \\
1mJy	 & 0 & 0 & 2 & 14 & 85 & 52 & 4 & 0 \\
100 µJy	 & 1 & 7 & 93 & 263 & 403 & 325 & 121 & 37 \\
10 µJy	 & 15 & 161 & 522 & 763 & 975 & 698 & 400 & 295 \\
1 µJy	 & 737 & 1218 & 1307 & 1307 & 1307 & 1296 & 1155 & 1054 \\
lg (background intensity) \\
($W m^{-2} sr^{-1}$) & -8.91 & -8.48 & -8.31 & -8.04 & -7.85 & -8.40 & -9.39 & -10.02
\end{tabular}
\end{minipage}
\end{table*}

\section{I-band Hubble diagram and luminosity function}
The I-band Hubble diagrams for the whole HDF galaxy sample is shown in Fig 4a,
together with the predicted curves for an Sbc galaxy with $M_I = -20$. 
This diagram is a natural extension to fainter magnitudes of the corresponding
distribution found by Crampton et al (1996) for the Canada-France survey to
I = 22.5 (their Fig 4). The corresponding Hubble diagram for the 327 galaxies
detected in all the four bands is shown in Fig 4b, with morphoplogical types
indicated by different symbols.  There is a clear trend towards fainter
magnitudes with increasing redshift, suggesting the photometric redshifts
are reasonable.  An even clearer trend emerges when only galaxies with
elliptical spectral-types are plotted (Fig 4c), as expected from the smaller
scatter in their absolute magnitudes (Fig 3).

The I-band completeness limit of the HDF is estimated as I = 27.5 mag. 
For the 1148 galaxies brighter than this limit, we have calculated the
luminosity function in three different redshift intervals (Fig 5). 
The luminosity function for $z < 1$ shows remarkably good consistency with
that found by Lilly et al (1996) for the Canada-France survey to I = 22.5 mag. 
They agree both in the normalisation and in the steep faint end slope, which
we find to be $\alpha = 1.5$ at $M_I > -22$, caused mainly by Sdm and HII 
galaxies. 
At  $M_I < -21$, there is clear evidence for strong evolution in the 
luminosity
function, with approximately a similar rate for the luminosity evolution 
as found for quasars (Boyle et al 1988) and starburst galaxies 
(Saunders et al 1990,
Rowan-Robinson et al 1993, Oliver et al 1995, Rowan-Robinson 1996). 
Although some optical galaxy redshift surveys have claimed to be inconsistent
with luminosity evolution (Broadhurst et al 1988, Colless et al 1990), recent
studies have begun to find hints of luminosity evolution in the galaxy
luminosity function (Colless 1995, Lilly et al 1996). A study of the
type-dependence of the luminosity function here shows that their faint end 
are mainly dominated by late-type spirals and HII galaxies (ie. galaxies
undergoing star formation and luminosity evolution). 

\section{Prediction of infrared and submillimatre fluxes}
It is a goal for many space projects and ground-based telescopes, existing or
planned in the infrared and submillimetre (eg ISO, SCUBA, FIRST, MMA/LSA), to
detect starburst galaxies at high redshift.  It is therefore of great interest
to predict the infrared and submillimetre fluxes for the HDF galaxies.

The very blue colours of most of the HDF galaxies suggest that they are
undergoing a strong episode of star formation, and this is supported by the
high incidence of peculiar and merging morphologies.  We assume that the SEDs
of these galaxies at infrared and submillimetre wavelengths can be well
represented by the standard starburst model of Rowan-Robinson and Efstathiou
(1993).  For the relative normalisation between far-infrared and optical,
we make an assumption similar to that made by Pearson and Rowan-Robinson
(1996), namely that  $F = \nu S_\nu (60\mu m )/\nu S_\nu (0.8 \mu m) = 10$ 
in the rest-frame. 
These authors claim that if the proportion of the total
bolometric power in a starburst, emerging in the optical-uv, is greater than
about $5-10\%$, then the optical galaxy counts at $B = 21-23$ mag. 
would be violated.

With these assumptions, we predict the flux for HDF galaxies at infrared or
submillimetre wavelength, allowing for their redshift and galaxy type. 
Table 1 gives the number of HDF galaxies expected brighter than a given
flux-density at wavelengths of interest to ISO, SCUBA, FIRST and proposed
large millimetre arrays (LSA/MMA).  
It appears that deep integrations by the ISO, SCUBA and FIRST are capable of 
detecting a number of HDF galaxies. 

We have also calculated the total background intensity expected from these
galaxies at each wavelength (see Table 1).  The resulting spectrum is similar
in shape and amplitude to that predicted by Pearson and Rowan-Robinson
(1996, see also Oliver et al 1992, Rowan-Robinson and Pearson 1996),
consistent with the picture that the majority of these galaxies are
strongly evolving starbursts, previously studied in far-infrared and 
sub-mJy radio surveys.

\section{Conclusions}
The conclusions of this study can be summarized as follows:
\begin{enumerate}
\item  A high proportion of HDF galaxies are undergoing a strong episode of
star formation, as evidenced by their very blue colours.  A wide range of
morphological types is found, with a high proportion of peculiar and merger
morphologies.
\item Fitting the multiband spectra with redshifted SEDs of galaxy types
E to HII, gives a plausible distribution of galaxy types, redshifts and
absolute magnitudes.  We find a median redshift of 1.6, with $68\%$ having
$z > 1$ and $31\%$ $z >2$.  The I-band absolute magnitude distributions as a
function of galaxy types show the expected trend of decreasing luminosity
towards later type galaxies. 
\item The derived I-band luminosity function agrees remarkably well with that
from the Canada-France survey (Lilly et al 1996) for $z < 1$ and shows strong
luminosity evolution at $M_I < -21$ for $1 < z < 3$, comparable to the rate seen
in quasars and starburst galaxies.
\item We have predicted the infrared and submillimetre fluxes assuming most 
of the
galaxies are undergoing a strong starburst.  Several planned space-borne and
ground-based deep surveys are capable of detecting interesting numbers of
HDF galaxies.
\end{enumerate}


\noindent{\bf Figure Captions}

\noindent {\bf Fig 1:}  (B-V) v. (U-B) for 327 HDF galaxies detected in all 4 bands. 
Predicted loci as a function of redshift, for $0 < z < 3$, are shown for 
E ($...$), Sbc ($-\ -\ -$) and HII ($---$) galaxies.

\noindent {\bf Fig 2:}  Redshift distribution as a function of galaxy spectral types.  
The peaks at $z= 2$ and 3 are likely artefacts of the photometric redshift 
method and our lack of knowledge of the far-UV spectra of galaxies. 
Distributions for the objects detected in all the 4 bands are plotted in dotted lines.

\noindent {\bf Fig 3:}  Absolute magnitude distribution as a function of galaxy spectral type. 
There is a plausible trend of decreasing peak luminosity towards later types.
Distributions for the objects detected in all the 4 bands are plotted in dotted lines.
 
\noindent {\bf Fig 4:} (a) I-band Hubble diagrams for the whole sample 
(1307 galaxies). The 9 galaxies with blue $V-I$ colours, detected only in I and V bands, 
are also plotted at a redshift of 3.57. These are candidates for high redshift objects. 
(b) I-band Hubble diagram for HDF galaxies detected in all 4 bands; 
E/SO ($\bullet$); Sa/Sab ($\star$); Sb/Sbc ($\ast$); Sc/Scd ($\circ$); Sd/Sdm ($\times$);
HII ($\Box$).  
(c) I-band Hubble diagram for elliptical galaxies.
The lines are the predicted relations for $M_I = -20$ mag. 

\noindent {\bf Fig 5:}  I-band luminosity function for 3 redshift ranges. 
Note the steep faint end slope ( $\alpha = 1.5$) caused mainly by Sdm and HII
galaxies and the strong luminosity evolution for $M_I < -21$ mag.
\vfil\break

\include{epsf}
\begin{figure*}
\caption{}
\epsfbox{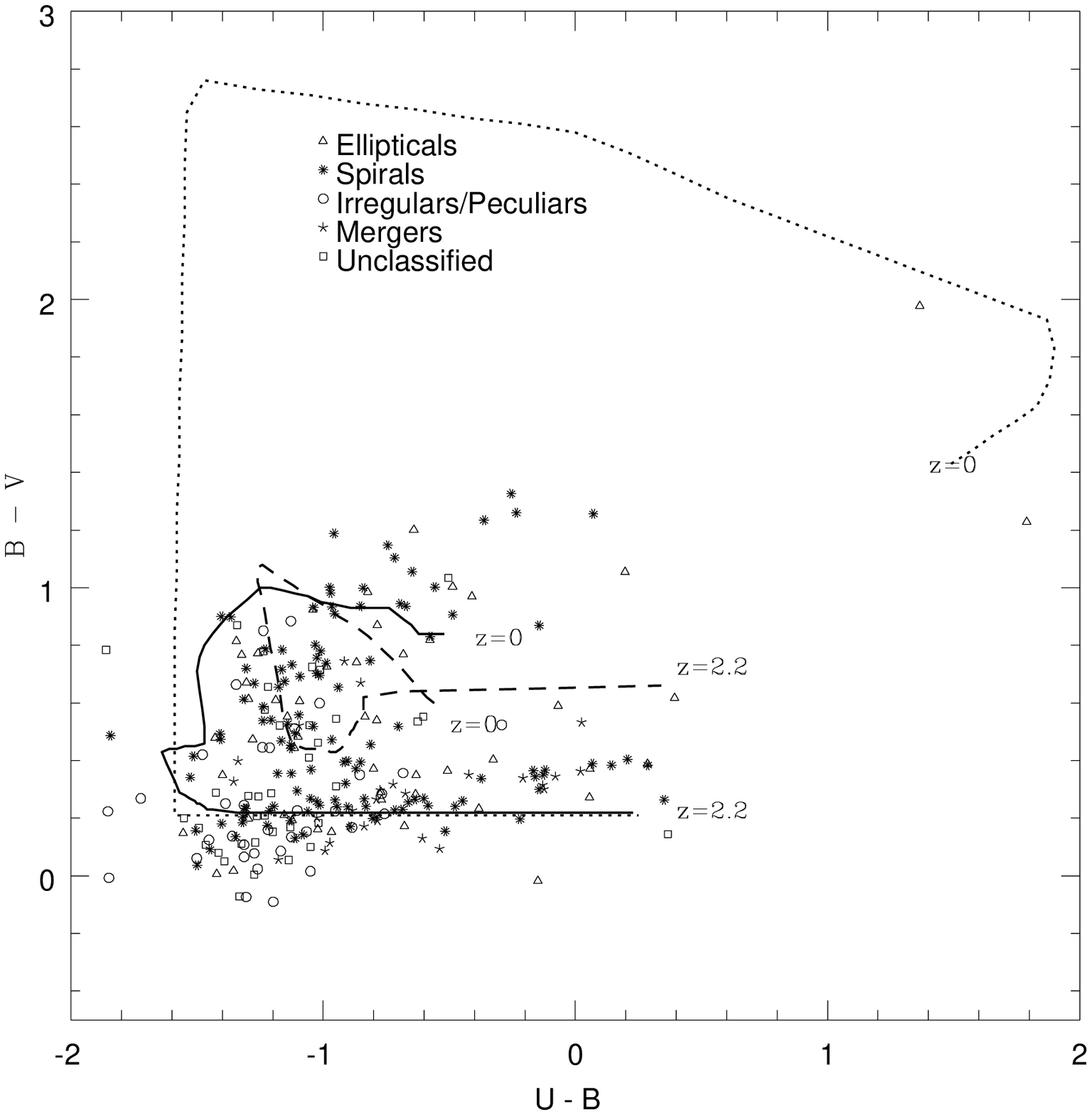}
\epsfverbosetrue
\end{figure*}
 
 \include{epsf}
\begin{figure*}
\caption{}
\epsfbox{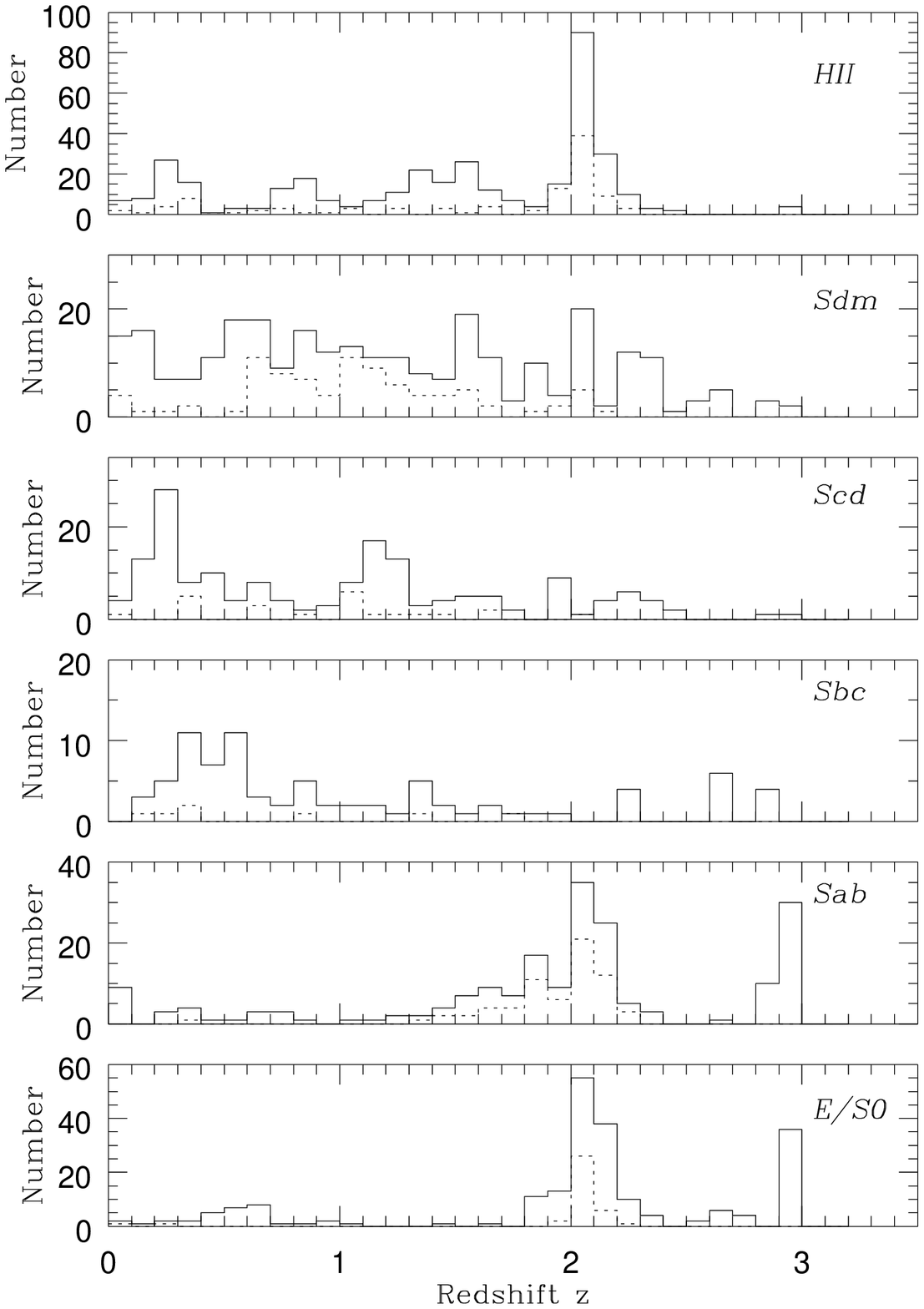}
\epsfverbosetrue
\end{figure*}
 
\include{epsf}
\begin{figure*}
\caption{}
\epsfbox{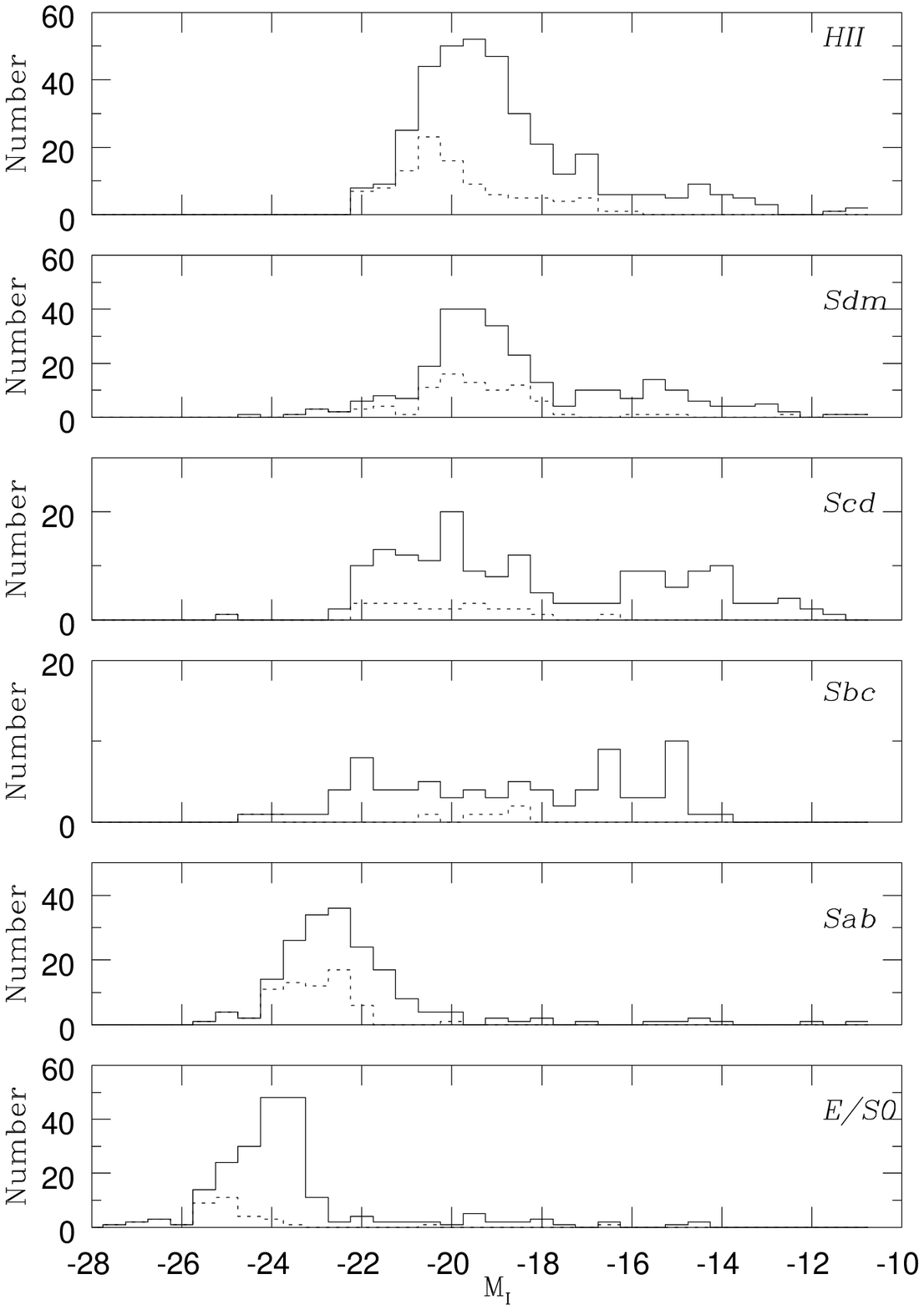}
\epsfverbosetrue
\end{figure*}
 
 \include{epsf}
\begin{figure*}
\caption{}
\epsfbox{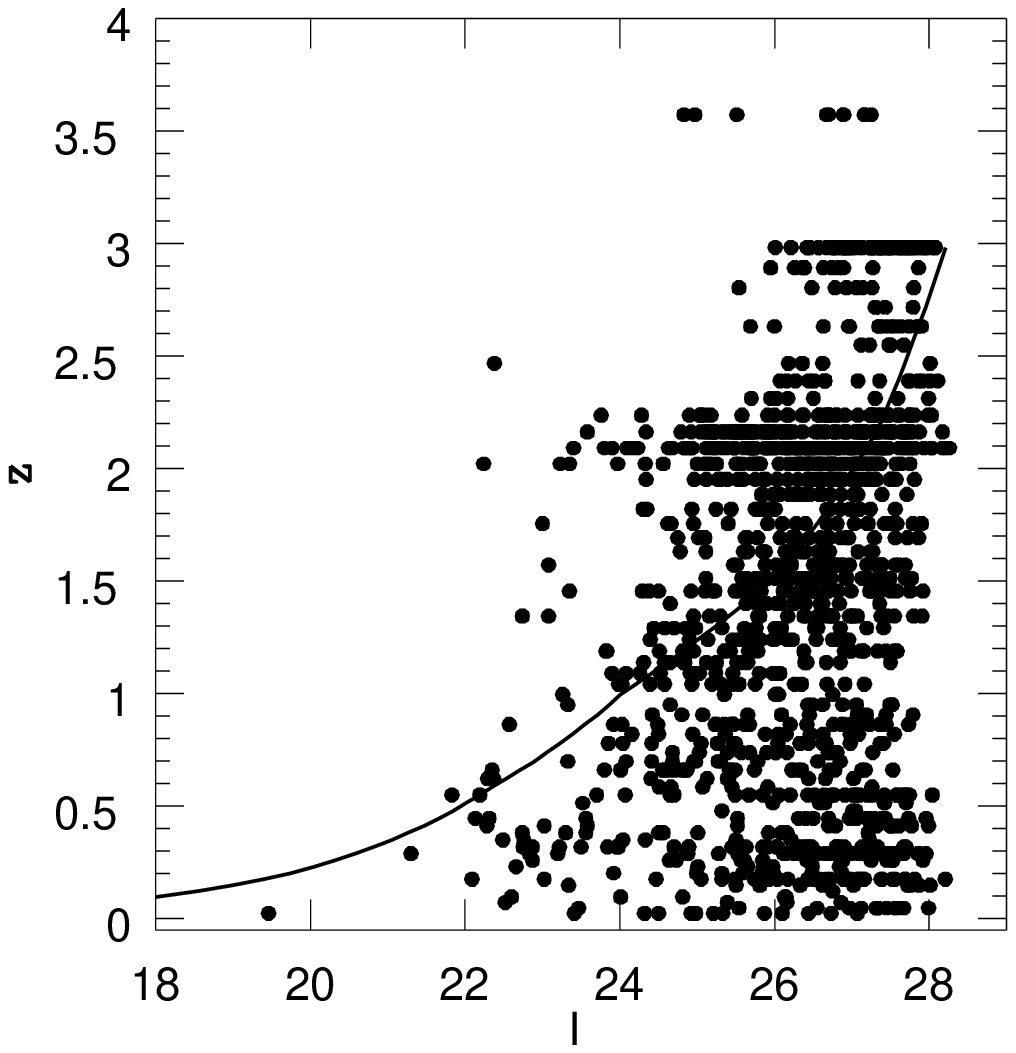}
\epsfverbosetrue
\end{figure*}
 
\include{epsf}
\begin{figure*}
\caption{}
\epsfbox{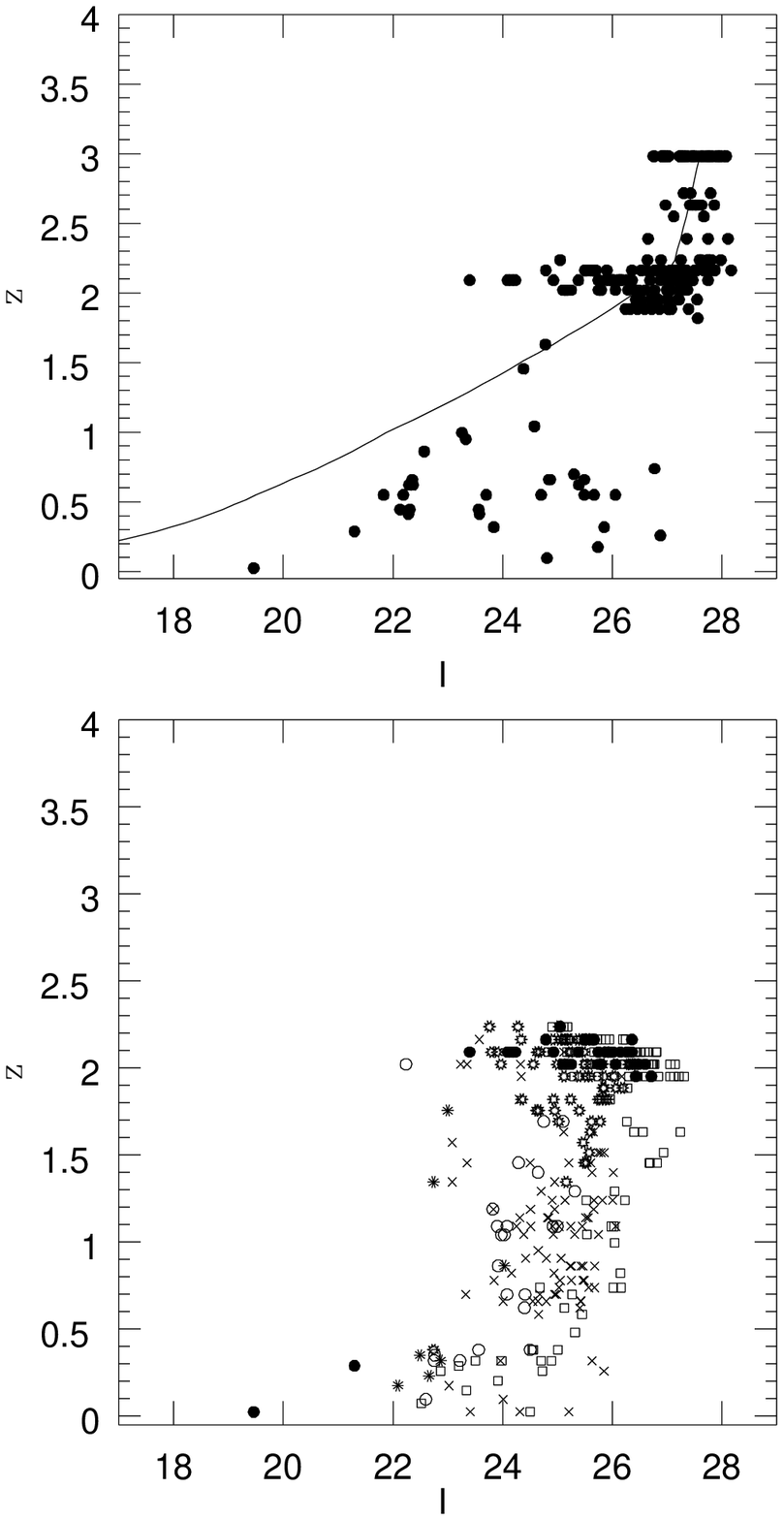}
\epsfverbosetrue
\end{figure*}
 
 \include{epsf}
\begin{figure*}
\caption{}
\epsfbox{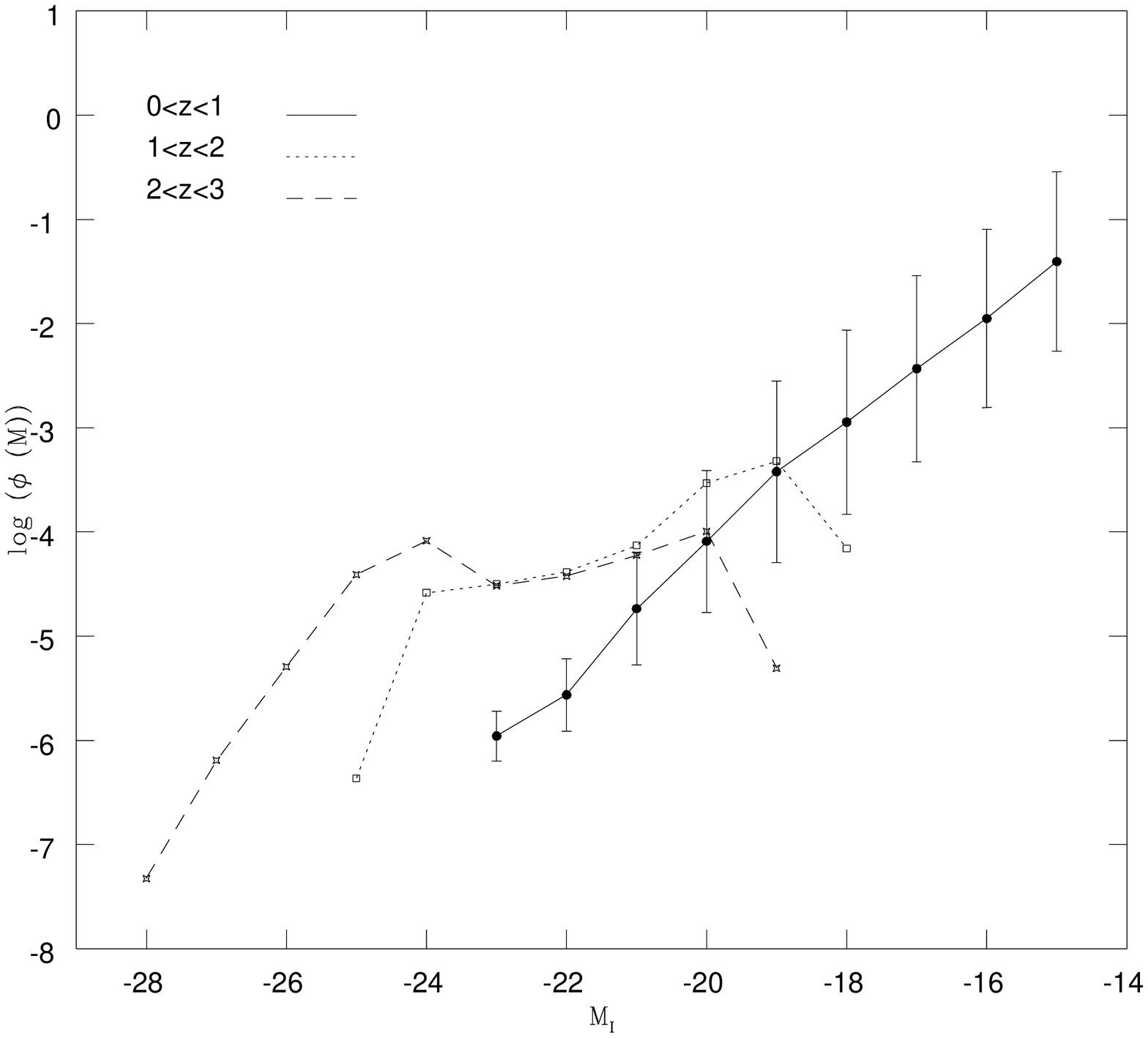}
\epsfverbosetrue
\end{figure*}
 
\end{document}